\documentclass[12pt,letterpaper]{article}

\usepackage{amsfonts}
\usepackage{amssymb}
\usepackage{epsfig}
\usepackage{amsmath}
\usepackage{graphicx}

\numberwithin{equation}{section}
\numberwithin{figure}{section}

\def\T{{\mathcal{T}}}
\newcommand{\pd}[2]{\frac{\partial#1}{\partial#2}}

\begin{document}

\title{Black Hole Statistical Mechanics and The Angular Velocity Ensemble}
\author{M. Thomson, \\ Department of Physics, University of Toronto, 60
Saint George Street,  \\ Toronto, ON, M5S 1A7, Canada
\\mitch.thomson@utoronto.ca \and C. C. Dyer \\ Physics and Astrophysics,
\\Department of Physical and Environmental Sciences,
\\ University of Toronto Scarborough, 1265 Military Trail, \\ Toronto, ON,
M1C 1A4, Canada \\dyer@astro.utoronto.ca}
\maketitle

\begin{abstract}
An new ensemble - the angular velocity ensemble - is derived using Jaynes' method of maximising entropy subject to prior information constraints. The relevance of the ensemble to black holes is motivated by a discussion of external parameters in statistical mechanics and their absence from the Hamiltonian of general relativity. It is shown how this leads to difficulty in deriving entropy as a function of state and recovering the first law of thermodynamics from the microcanonical and canonical ensembles applied to black holes.
\end{abstract}

\section{Introduction}

Jaynes' information theory approach \cite{Jaynes} to the foundations of statistical mechanics was influential in Bekenstein's linking of black hole entropy to horizon area \cite{Bekenstein}. Here we introduce a new ensemble - the angular velocity ensemble - derived using Jaynes' principle of maximising entropy subject to prior information constraints. The relevance of the ensemble is motivated by a discussion of black hole entropy as a function of state. In the statistical mechanics of an ideal gas the volume dependence of the entropy arises from an external field in the Hamiltonian. This is contrasted with the lack of an external parameter corresponding to angular momentum or angular velocity in the Hamiltonian of general relativity. The resulting difficulty in reproducing the entropy as a function of state and the first law of black hole mechanics from the microcanonical and canonical ensemble holes is highlighted. A brief review of the foundations of statistical mechanics is given to explain that the Jaynes' approach removes the central role of the microcanonical ensemble and therefore the rotational (or grand canonical) and angular velocity ensembles may still be applied in its absence.

\section{The Relationship of Entropy and Horizon Area}

The first law of black hole mechanics for a non-charged black hole of mass $M$, angular velocity $\Omega$, angular momentum $J$ and surface gravity $\kappa$ is

\begin{equation}\label{First Law BH mech}
\delta M = \frac{\kappa}{8\pi}\delta A + \Omega \delta J.
\end{equation}

\noindent In the derivation of this law from neighbouring solutions to Einstein's field equations $A$ is the physical area of the horizon \cite{Bardeen_et_al}. Mathematically, black hole entropy could be related to any function of $A$ to result in an equation in the form of the first law of thermodynamics \cite{Bekenstein}. Physically, Hawking's discovery \cite{Hawking} of the black hole temperature

\begin{equation}\label{black hole temperature}
T=\frac{\kappa}{2\pi},
\end{equation}

\noindent fixes the relation between black hole horizon area and entropy as

\begin{equation}\label{Entropy - area relation}
\mathcal{S}=\frac{A}{4}.
\end{equation}

\noindent The central role of horizon area originates from its presence in equation (\ref{First Law BH mech}) and leads to claims that black hole entropy is a physical property of the horizon \cite{Sorkin}. It is therefore revealing to derive black hole entropy, without reference to the horizon area, by integrating the first law of thermodynamics.

The task is to find the function of state $\mathcal{S}(M,J)$ such that:

\begin{equation}
d\mathcal{S}=\frac{dM}{T}-\frac{\Omega}{T}dJ.
\end{equation}

\noindent To do this consider the change in entropy when an initial state $(M_1,J_1)$ is transformed adiabatically into a final state $(M_2,J_2)$

\begin{equation}
\mathcal{S}_2-\mathcal{S}_1=\displaystyle\int_1^2d\mathcal{S}=\displaystyle\int_1^2 \frac{1}{T}dM-\displaystyle\int_1^2\frac{\Omega}{T}dJ.
\end{equation}

\noindent As $d\mathcal{S}$ is an exact differential the integration path can be freely chosen. So first hold $J$ constant and change $M_1$ to $M_2$, arriving at an intermediate state $(M_2,J_1)$, and then hold $M$ constant whilst $J_1$ is changed to $J_2$ to arrive at the final state. Labeling the intermediate state as 3, we have

\begin{equation}
\mathcal{S}_2-\mathcal{S}_1=\displaystyle\int_1^3 \frac{1}{T}dM-\displaystyle\int_3^2\frac{\Omega}{T}dJ.
\end{equation}

\noindent Angular velocity $\Omega$ is given as a function of state variables $M$ and $J$ by the Smarr relation 

\begin{equation}\label{Smarr Omega}
\Omega=\frac{J}{2M(M^2+(M^4-J^2)^{1/2})}.
\end{equation}

\noindent The Hawking temperature, equation (\ref{black hole temperature}), is obtained as a function of state using the Smarr relation for the surface gravity $\kappa$

\begin{equation}
\kappa=\frac{(M^4-J^2)^{1/2}}{2M(M^2+(M^4-J^2)^{1/2}}.
\end{equation}

\noindent Inserting these into the entropy integral gives

\begin{equation}
\mathcal{S}_2-\mathcal{S}_1=2\pi\displaystyle\int_{M_1}^{M_2} \frac{2M(M^2+\sqrt{M^4-J_1^2})}{\sqrt{M^4-J_1^2}}dM-2\pi\displaystyle\int_{J_1}^{J_2}\frac{J}{\sqrt{M_2^4-J^2}}dJ,
\end{equation}

\noindent and performing the integral leads to

\begin{equation}
\mathcal{S}_2-\mathcal{S}_1=2\pi\left(M_2^2+\sqrt{M_2^4-J_2^2}-M_1^2-\sqrt{M_1^4-J_1^2}\right).
\end{equation}

\noindent It is clear from this expression that the entropy must be

\begin{equation}
\mathcal{S}(M,J)=2\pi\left(M^2+\sqrt{M^4-J^2}\right) + \ constant
\end{equation}

\noindent From the Smarr relation for area as a function of state

\begin{equation}\label{Horizon area}
A=8\pi\left(M^2+\sqrt{M^4-J^2}\right),
\end{equation}

\noindent it is clear that this entropy expression is in agreement with equation (\ref{Entropy - area relation}).

The entropy of a black hole can therefore be derived without recourse to the area of the horizon. Evidently, entropy is 
proportional to the same function of $M$ and $J$ as the area, but this fact has played no logical part in the above 
derivation. This analysis shows there is no logical necessity for black hole entropy to be physically related to 
the horizon area. 

Black hole entropy calculations often result in entropy of the form

\begin{equation}
\mathcal{S}=\frac{1}{4}\int_{\mathcal{H}} dA,
\end{equation}

\noindent where the integral is over the horizon $\mathcal{H}$ of a specific solution selected from the Kerr family. Care must then be taken in substituting the Smarr relation, equation (\ref{Horizon area}), and claiming that the entropy has been derived as a function of state. After all, the Smarr relation for area can not be proved by evaluating the horizon area of a single metric.

\section{The Applicability of the Microcanonical and Canonical Ensembles}\label{The Applicability of the Microcanonical Ensemble to Gravity}

The particles of a classical ideal gas are confined to a container of volume $V$ by an external potential field $U(q_i)$ satisfying

\begin{equation}
U(q_i)= \left\{
\begin{aligned}
&0\ \ &q_i \in V\\
&\infty \ &q_i \notin V
\end{aligned}
\right . ,
\end{equation}

\noindent where $q_i$ is the position of the $i^{th}$ particle. In addition to being dependent on the phase space variables $q_i$ and momentum $p_i$ of each particle $i$, the Hamiltonian of the system is dependent upon $V$ as an external parameter:

\begin{equation}
H=\displaystyle\sum_{i=1}\left(\frac{p_i^2}{2m}+U(q_i)\right).
\end{equation}

\noindent Using the shorthand notation $(q,p)$ for all phase space coordinates this dependency can be expressed as $H=H(q,p;V)$. It is through the Hamiltonian that the volume dependency of the microcanonical ensemble phase space density arises

\begin{equation}\label{microcanonical phase}
\rho_M(q,p;E,V)=\delta(E-H(q,p;V)).
\end{equation}

\noindent The entropy

\begin{equation}\label{stat mech entropy}
\mathcal{S}(E,V)=-k_B\int dp dq \ \rho_M(p,q;E,V)\log \rho_M(p,q;E,V)
\end{equation}

\noindent is a function of two state variables $(E,V)$ which is appropriate in the case of fixed particle number.

Similarly, the volume dependence of the canonical ensemble

\begin{equation}\label{canonical phase}
\rho_C(q,p;T,V)=\frac{e^\frac{-H(q,p;V)}{k_BT}}{Z(T,V)},
\end{equation}

\noindent where $Z(T,V)$ is the canonical partition function

\begin{equation}\label{canonical partition function}
Z_C(T,V)=\int dq dp \ e^\frac{-H(q,p;V)}{k_BT},
\end{equation}

\noindent also arises through the Hamiltonian. In both cases the entropy is dependent on $V$ because it is an external parameter in the Hamiltonian.

The Hamiltonian of General Relativity for a $d$ dimensional spacetime $\mathcal{M}$ with a Lorentzian boundary $\mathcal{T}$ is

\begin{equation}
H= \int_{\mathcal{M}} d^dx\sqrt{-g}\left[N\mathcal{H}+V^{i}\mathcal{H}_i\right]+ 
\int_\T d^{d-1}x\sqrt{\sigma}\left[N\varepsilon-V^ij_i\right],
\end{equation}

\noindent where $\mathcal{H}$ is the Hamiltonian constraint, $\mathcal{H}_i$ is the momentum constraint, $N$ and $V^i$
are the lapse and shift, and $\varepsilon$ and $j_i$ are the energy and momentum surface densities 
\cite{Brown_York_quasilocal_energy}.  The angular velocity $\Omega$ is entirely absent and although the angular momentum
 density $j_i$ does appear, it is a function of phase space variables - it is not an external parameter. Denoting the 
physical degrees of freedom of the gravitational field as $\gamma$ and their canonical momenta as $\pi$, a heuristic microcanonical phase space density is

\begin{equation}\label{microcanonical phase space density}
\rho_M(\gamma,\pi)=\delta(M-H(\gamma,\pi)).
\end{equation}

\noindent Using a functional integral to integrate over phase space, the entropy would be

\begin{equation}\label{black hole entropy from microcanonical phase density}
\mathcal{S}(M) = -k_B\int D\gamma D\pi \ \rho_M(\gamma,\pi) \log \rho_M(\gamma,\pi).
\end{equation}

\noindent The problem is that all values of $j_i$ are integrated over and entropy is a function of $M$ only. There is consequently no way of determining the entropy as a function of two state variables using the microcanonical ensemble. 

The microcanonical ensemble was applied to black holes by Brown and York \cite{Brown_York_microcan_action}. Although they did not relate it to a missing external parameter, Brown and York did encounter this problem on route to their definition of the microcanonical density of states. They were forced to introduce an angular momentum dependence by fixing $j_i$ as a boundary condition in the path integral density of states. This was motivated by comparing the nature of time translations in non-gravitational physics, where energy is the value of the Hamiltonian that generates unit time translations, and general relativity, where time is ``many fingered'', and therefore both $\varepsilon$ and $j_i$ play a role in generating unit translations of the boundary. The real reason $j_i$ needs to be fixed in the microcanonical ensemble, however, is because it is not an external parameter.

By arbitrarily fixing $j_i$, and therefore its surface integral $J$, it is unclear how the gravitational microcanonical ensemble should be related to a standard microcanonical phase space density in the form of equation (\ref{microcanonical phase space density}). Is the integral over phase space in equation (\ref{black hole entropy from microcanonical phase density}) to be restricted to phase points with the specified value of $J$? If so, how is a small change in $J$ to be implemented in order to recover the first law?

In the case of the ideal gas, the first law is recovered by variation of the external parameter $V$. 
Consider the canonical ensemble (\ref{canonical phase}) substituted in to the entropy (\ref{stat mech entropy}):

\begin{equation}
\mathcal{S}= k_B\beta\left\langle E\right\rangle + k_B \log Z_C , 
\end{equation}

\noindent where $\beta=\frac{1}{k_B T}$ and the notation $\left\langle E\right\rangle$ denotes the phase space average of $E$:

\begin{equation}
\left\langle E\right\rangle =\int dqdp \ E(q,p) \rho_C(q,p).
\end{equation}

\noindent As a function of $V$ and $\beta$ the entropy's differential is

\begin{equation}
d\mathcal{S} = k_B d\beta \left\langle E\right\rangle +  k_B\beta d\left\langle E\right\rangle + k_B \pd{}{V}\log Z_C dV + k_B \pd{}{\beta}\log Z_C d\beta.
\end{equation}

\noindent The derivatives of the partition function can be obtained from its definition (\ref{canonical partition function}):

\begin{equation}
\pd{}{V}\log Z_C(\beta,V) = -\beta \left\langle \pd{E}{V}\right\rangle,
\end{equation}

\begin{equation}
\pd{}{\beta}\log Z_C(\beta,V) = -\left\langle E\right\rangle,
\end{equation}

\noindent and therefore $d\mathcal{S}$ is

\begin{equation}
d\mathcal{S} = k_B\beta d\left\langle E\right\rangle - k_B \beta  \left\langle \pd{E}{V}\right\rangle dV.
\end{equation}

\noindent The first law of thermodynamics

\begin{equation}
d\left\langle E\right\rangle =Td\mathcal{S}+\left\langle P\right\rangle dV,
\end{equation}

\noindent is recovered because $P = \pd{E}{V}$ is the definition of pressure. The key point is that the first law has only been recovered from the canonical ensemble because one of the thermodynamic state variables was an external parameter in the Hamiltonian, which is not the case in general relativity.

\section{Ergodicity, Maximum Entropy Principle and the Foundations of Statistical Mechanics}

The foremost concern in the foundations of statistical mechanics is to explain why thermodynamical state variables can be related to ensemble averages so successfully. Consider a classical system, with phase space coordinates $(q,p)$, upon which a measurement is made of a thermodynamical property $\chi$ that is uniquely determinable from the actual microstate of the system. Experimentally, such a measurement will typically be a short time average. When the measurement begins at time $t_0$ the system is in state $(q(t_0),p(t_0))$, and during the measurement it follows a path through phase space determined by the equations of motion. When the measurement concludes after an elapsed time $t_{obs}$ the state of the system is $(q(t_0+t_{obs}),p(t_0+t_{obs}))$. In principle, the result of the measurement is

\begin{equation}\label{chi obs}
\chi_{obs}=\frac{1}{t_{obs}}\displaystyle\int_{t_0}^{t_0+t_{obs}}\chi\left(q(t),p(t)\right)dt.
\end{equation}

\noindent In reality, thermal systems have so many degrees of freedom that it is impossible either to determine the initial state or to solve the equations of motion. Instead, statistical mechanics relates the result to a phase space average

\begin{equation}
\langle \chi \rangle =\displaystyle\int\chi(q,p)\rho(q,p)dq dp \ ,
\end{equation}

\noindent for an appropriately chosen ensemble $\rho(q,p)$. The key issue is to explain the relationship between $\chi_{obs}$ and $\langle \chi \rangle$. The ergodic hypothesis and equal a priori probabilities are the two most well known approaches \cite{Tolman_stat_book}.

The ergodic hypothesis asserts that an isolated system passes in succession through each point of phase space compatible with the system's energy before eventually returning to its starting point. Accepting this assertion as true, it can be shown that $\chi_{obs}=\langle \chi \rangle$, and therefore the equality of time and ensemble averages is a result of standard dynamics. The ergodic hypothesis fails as a rationalisation of statistical mechanics because the time-period over which a system is judged to be ergodic is usually infinity. But for  $\chi_{obs}$ to equal $\langle \chi \rangle$ the system must be ergodic over a time of order $t_{obs}$, and if this were true of all isolated systems it would have been impossible to experimentally distinguish equilibrium states from non-equilibrium states \cite{Jaynes_where_do_we_stand}.

The principle of equal a priori probabilities asserts that all microstates of energy $E$ should be given equal probability in the ensemble representing an isolated system of energy $E$. It therefore places the microcanonical ensemble at the centre of the justification of statistical mechanics. The canonical ensemble is then obtained by applying the microcanonical ensemble to a system consisting of subsystems exchanging heat and showing the subsystems form a canonical ensemble \cite{Hill}. Interestingly, the subsystems are required to be weakly interacting over short distances only, so gravity would be difficult to include in this framework. Jaynes' information theory approach to the rationalisation of statistical mechanics \cite{Jaynes} is an extension of the principle of a priori equal probabilities to situations where more than one piece of information is available. Jaynes' criterion states an ensemble should be the assignment that agrees with the known information whilst being unbiased in all other respects. Claude Shannon, in the context of information theory, shows that maximising entropy, subject to the known information constraints, leads to this assignment \cite{Shannon}.

\section{The Rotational Ensemble}\label{The Rotational Ensemble}

The rotational ensemble was first defined by Gibbs \cite{Gibbs} and referred to as the grand canonical ensemble by Gibbons and Hawking \cite{Gibbons_Hawking_Action_Integrals}. The ensemble can be derived by the Jaynes' procedure \cite{Jaynes_Heims}.

The grand canonical ensemble for an ideal gas, which is a function of the chemical potential $\mu$, is a Legendre transformation of the microcanonical ensemble, which is a function of the number of particles $N$. In contrast, the rotational ensemble is not a Legendre transformation of a microcanonical ensemble because the microcanonical ensemble is neither a function of $J$ nor $\Omega$.

\noindent The rotational ensemble is

\begin{equation}\label{rotational ensemble}
\rho(q,p)=\frac{e^{-\beta E(q,p)-\lambda J(q,p)}}{Z(\beta,\lambda)}.
\end{equation}

\noindent where 

\begin{equation}
\beta=\frac{1}{k_BT},
\end{equation}

\begin{equation}
\lambda=\frac{-\Omega}{k_BT},
\end{equation}

\noindent and $Z(\beta,\lambda)$ is the partition function

\begin{equation}
Z(\beta,\lambda)= \int dqdpe^{-\beta E(q,p)-\lambda J(q,p)}.
\end{equation}

\noindent Substituting the phase space density (\ref{rotational ensemble}) into the entropy (\ref{stat mech entropy}) gives

\begin{align}\label{rotational entropy}
\mathcal{S}=k_B\beta \left\langle E\right\rangle +k_B\lambda \left\langle J\right\rangle +k_B\log Z(\beta,\lambda).
\end{align}

\noindent In Jaynes' approach the Lagrange multipliers $\lambda$ and $\beta$ are not independent variables. They are dependent on the prior information $\left\langle M\right\rangle $ and $\left\langle J\right\rangle $ through the equations:

\begin{equation}
-\pd{\log Z(\beta,\lambda)}{\beta}=\left\langle E\right\rangle ,
\end{equation}

\begin{equation}
-\pd{\log Z(\beta,\lambda)}{\lambda}=\left\langle J\right\rangle .
\end{equation}

\noindent However, finding $d\mathcal{S}$ is helped by reversing the logical dependency and considering $\left\langle M\right\rangle $ and $\left\langle J\right\rangle $ to be functions of $\lambda$ and $\beta$. This leads to the differential relationship

\begin{equation}\label{temp rot dS}
d\mathcal{S}=k_B\beta d\left\langle E\right\rangle +k_B\lambda d\left\langle J\right\rangle ,
\end{equation}

\noindent which is the first law of thermodynamics

\begin{equation}
d\left\langle E\right\rangle = T d\mathcal{S} +\Omega d\left\langle J\right\rangle.
\end{equation}

\noindent The rotational ensemble is therefore capable of reproducing the first law and leads to an entropy that is a function of two state variables: $T$ and $\Omega$. The dependency on $\Omega$ arises not as an external field in the Hamiltonian but as a parameter in the definition of the ensemble. 

The phase space over which the rotational ensemble is defined includes solutions with all values of $M$ and $J$. To calculate the entropy of the Schwarzschild solution, one would specify $\left\langle J\right\rangle =0$, and then calculate the entropy from equation (\ref{rotational entropy}). Even at the classical level there will be more than just the Schwarzschild solution contributing to the ensemble averages. This process is very different from looking for an ensemble to ``represent'' the Schwarzschild solution.

\section{Landau's Canonical Rotational Ensemble}

The difficulty of using the microcanonical and canonical ensemble in the absence of an external field in the Hamiltonian is not specific to black hole statistical mechanics; it is equally relevant to the case of a rotating system in non-gravitational classical mechanics. Landau's application \cite{Landau1} of the canonical ensemble to a classical body in uniform rotation about an axis is therefore considered in this section and shown to be equivalent to the rotational ensemble.

The relationship between the energy $E_{rot}(p,q)$ of a body seen from a co-rotating frame of reference and the energy $E(p,q)$ in an inertial frame of reference is

\begin{equation}\label{H_rot}
E_{rot}(p,q)=E(p,q)-\Omega J(p,q).
\end{equation}

\noindent The angular velocity $\Omega$ is therefore an external parameter of $E_{rot}(p,q;\Omega)$ and

\begin{equation}
\pd{E_{rot}(p,q;\Omega)}{\Omega}=-J(p,q).
\end{equation}

\noindent A canonical ensemble based on $E_{rot}(p,q;\Omega)$ and the partition function

\begin{equation}\label{Landau Z rot}
Z(T,\Omega)=\int dqdp\ e^{-\frac{E_{rot}(p,q;\Omega)}{k_BT}},
\end{equation}

\noindent leads to a form of the first law for the rotating observer

\begin{equation}\label{temp rotation first law}
d\left\langle E_{rot}\right\rangle =Td\mathcal{S}- \left\langle J\right\rangle d\Omega.
\end{equation}

\noindent This must now be converted to the first law relevant to the fixed coordinate system. From equation (\ref{H_rot}), $\left\langle E_{rot}\right\rangle =\left\langle E\right\rangle -\Omega\left\langle J\right\rangle$ and therefore

\begin{equation}
d\left\langle E_{rot}\right\rangle=d\left\langle E\right\rangle-d\Omega \left\langle J\right\rangle-\Omega
d\left\langle J\right\rangle.
\end{equation}

\noindent Inserting this into (\ref{temp rotation first law}) leads to the first law

\begin{equation}
d\left\langle E\right\rangle =Td\mathcal{S}+ \Omega d\left\langle J\right\rangle.
\end{equation}

It therefore appears that, for non-gravitational rotating systems, $\Omega$ can be turned into a parameter in the energy and a canonical ensemble analysis used. However, if $E_{rot}(p,q;\Omega)$ is substituted from (\ref{H_rot}) into the canonical ensemble partition function (\ref{Landau Z rot}), it is seen to be the partition function of the rotational ensemble

\begin{equation}
Z(T,\Omega)=\int dqdp\ e^{-\frac{E(p,q)}{k_BT}+\frac{\Omega J(p,q)}{k_BT}}.
\end{equation}

\noindent This method does not therefore offer an independent method of applying the canonical ensemble to rotating systems and recovering the first law. In any case, the relationship (\ref{H_rot}) between energies in rotating and inertial coordinates is not applicable to the mass of black holes; a rotating Kerr black hole viewed by a co-rotating observer is not a Schwarzschild black hole owing to the differential rotation.

\section{The Angular Momentum Ensemble}\label{The Angular Momentum Ensemble}

In the rotational ensemble the prior information about the thermal state of the system is the mass $M$ (or the energy $E$) and angular momentum $J$. Here we derive an ensemble that is applicable when angular velocity $\Omega$ is the prior information in place of $J$. 

Label the possible states of a black hole, including states with all possible values of $J$ and $M$, by $x_i$. Because $\Omega$ is an external parameter, and not a physical property of the states $x_i$, it must be added as a coordinate to the phase space in order for the state of the system to be fully specified. In other words, microstates of the system are specified by the pair $(x_i,\Omega)$. In addition to $\Omega$ another piece of information must be provided. It cannot be the mass $M(x_i)$ as it is independent of $\Omega$, and a function on the augmented phase space is required. A natural choice is to define a rotational equivalent of enthalpy:

\begin{equation}
\textbf{H}(x_i,\Omega)= M(x_i) - \Omega J(x_i).
\end{equation}

A statistical ensemble $p(x_i,\Omega)$ is an assignment of a probability to each state $(x_i,\Omega)$. For the ensemble to be representative, the ensemble averages $\left\langle \textbf{H}\right\rangle$ and $\left\langle \Omega\right\rangle$ must equal the known values. The as yet undetermined probability assignment $p(x_i,\Omega)$ must therefore satisfy the constraints:

\begin{equation}\label{constraint 1}
\int_0^\infty d\Omega \sum_{x_i}\ p(x_i,\Omega)\textbf{H}(x_i,\Omega) = \left\langle \textbf{H}\right\rangle ,
\end{equation}

\begin{equation}\label{constraint 2}
\int_0^\infty d\Omega \sum_{x_i}\ p(x_i,\Omega)\Omega(x_i,\Omega) = \left\langle \Omega\right\rangle ,
\end{equation}

\noindent where $\Omega(x_i,\Omega)\equiv \Omega$. These constraints are in addition to the standard rule of probability:

\begin{equation}\label{constraint 3}
\int_0^\infty d\Omega\sum_{x_i} p(x_i,\Omega) =1.
\end{equation}

\noindent Of the many possible assignments $p(x_i,\Omega)$ that satisfy these constraints, Jaynes' criterion \cite{Jaynes} selects the assignment that is unbiased in all other respects by maximising the entropy

\begin{equation}
\mathcal{S}=-k_B\int_0^\infty d\Omega \sum_{x_i}\ p(x_i,\Omega)\log p(x_i,\Omega),
\end{equation}

The entropy is maximised subject to the constraints by introducing Lagrange multipliers $\nu$, $\beta$ and $\mu$ and solving

\begin{multline}
\pd{\mathcal{S}}{p(x_j,\Omega')}
+\beta \pd{}{p(x_j,\Omega')}\left(\left\langle \textbf{H}\right\rangle -\displaystyle\int_0^\infty d\Omega\sum_{x_i}p(x_i,\Omega)\textbf{H}(x_i,\Omega)\right) \\
+\nu \pd{}{p(x_j,\Omega')}\left(\left\langle \Omega \right\rangle -\displaystyle\int_0^\infty d\Omega\sum_{x_i}p(x_i,\Omega)\Omega(x_i,\Omega)\right) \\
+ \mu \pd{}{p(x_j,\Omega')}\left(1-\displaystyle\int_0^\infty d\Omega\sum_{x_i}p(x_i,\Omega)\right)=0.
\end{multline}

\noindent This leads to

\begin{equation}
p(x_i,\Omega)=e^{-\mu}e^{-\beta \textbf{H}(x_i,\Omega)-\nu \Omega(x_i,\Omega)},
\end{equation}

\noindent where a factor of $e$ has been absorbed into $e^{-\mu}$. The Lagrange multiplier $\mu$ can be determined from the normalisation constraint (\ref{constraint 3})

\begin{equation}
\mu=\log \left(\int_0^\infty d\Omega \sum_{x_i}\  e^{-\beta H(x_i,\Omega)-\nu \Omega(x_i,\Omega)}\right)\equiv \log Z(\beta,\nu),
\end{equation}

\noindent where the angular momentum partition function is defined by the second step. The ensemble that satisfies the constraints and is unbiased in all other respects is therefore

\begin{equation}\label{Angular Momentum Ensemble}
p(x_i,\Omega)=\frac{e^{-\beta \textbf{H}(x_i)-\nu \Omega(x_i,\Omega)}}{Z(\beta,\nu)}.
\end{equation}

\noindent Further information on the Lagrange multipliers is obtained by comparison with the first law of thermodynamics. With the ensemble probability (\ref{Angular Momentum Ensemble}) substituted, the entropy (\ref{stat mech entropy}) is

\begin{align}
\mathcal{S}=k_B\beta \left\langle \textbf{H}\right\rangle +k_B\nu \left\langle \Omega\right\rangle +k_B\log Z(\beta,\nu).
\end{align}

\noindent The differential is

\begin{equation}\label{Temp dS}
d\mathcal{S}=k_B\beta d\left\langle \textbf{H}\right\rangle +k_B\nu d\left\langle \Omega\right\rangle ,
\end{equation}

\noindent where use has been made of

\begin{equation}\label{H=dlogZ/dbeta}
\left\langle \textbf{H}\right\rangle =-\pd{\log Z(\beta,\nu)}{\beta},
\end{equation}

\begin{equation}\label{Omega=dlogZ/dlambda}
\left\langle \Omega\right\rangle =-\pd{\log Z(\beta,\nu)}{\nu}.
\end{equation}

\noindent The first law of black hole mechanics with the mass replaced by enthalpy is

\begin{equation}
d\textbf{H} = Td\mathcal{S}-Jd\Omega.
\end{equation}

\noindent By comparison with (\ref{Temp dS}) the Lagrange multipliers are seen to be

\begin{equation}
\beta=\frac{1}{k_BT},
\end{equation}

\begin{equation}
\nu=\frac{J}{k_BT}.
\end{equation}

\noindent and the angular velocity ensemble is

\begin{equation}
p(x_i,\Omega)=\frac{exp\left(-\frac{\textbf{H}(x_i,\Omega)}{k_BT}-\frac{J\Omega}{k_BT}\right)}{Z(T,J)}.
\end{equation}

\noindent It may appear that substituting $\textbf{H}(x_i,\Omega)$ will convert the angular velocity ensemble back into the rotational. However, in this case $J$ is a parameter determined by the constraints, not a function of phase space as in the case of the rotational ensemble. The angular velocity ensemble is equally relevant to rotating systems in non-gravitational classical statistical mechanics.

The angular velocity ensemble is mathematically identical to the pressure ensemble defined by M\o{}ller [unpublished] (see \cite{Ter_Haar:1971xt}). M\o{}ller derived the pressure ensemble by considering an ideal gas in a canister having a movable piston with a weight on top. In this situation the pressure is a fixed external parameter (although not one that appears in the Hamiltonian) and the volume must be specified, along with the phase space point, to determine the state of the system. The pressure ensemble was re-derived by Jaynes as a demonstration of the information theory approach to statistical mechanics \cite{Jaynes}. The Jaynesian approach is crucial in the derivation of the angular velocity ensemble where recourse can not be made to a simple physical model such as the ideal gas in the container.

\section{Discussion}

The angular velocity ensemble has been derived here by Jaynes' information theory approach to statistical mechanics. Both the rotational and angular velocity ensembles lead to an entropy that is a function of state and  are able to reproduce the first law of thermodynamics in the absence of an external field in the Hamiltonian corresponding to angular velocity or momentum. The discussion here has focused on the Hamiltonian of general relativity but leads to the more general point that the dynamics of the system must involve an external parameter in the Hamiltonian if the microcanonical or canonical ensemble are to be used in black hole statistical mechanics.

The path integral is a tool to calculate the canonical partition function, with other partition functions subsequently obtained by Legendre transformations. It works in situations where the thermal state can be determined by the temperature and external parameters in the Hamiltonian. Application of the path integral to various ensembles in black hole statistical mechanics involves fixing additional dynamical data and therefore the direct link between the path integral and the partition function is broken \cite{Brown_etal_td_ensembles}. From a Jaynesian perspective the reasoning of statistical mechanics is extremely robust, so rather than elevating the path integral to a fundamental position, we are of the opinion that the partition functions are fundamental irrespective of the ability of path integrals to evaluate them. 

The derivation of black hole entropy from an integration of the first law is logically dependent on two points: a black hole is a system that returns to an equilibrium state described by the Smarr relations; and its temperature is given by the Hawking temperature. Any loss of information behind the horizon is not crucial. Focusing on black hole entropy as a function of state, and not a property of the horizon, removes confusion as to why the horizon should have microstates when it is locally the same as anywhere else in the vacuum. It also, however, may focus on other foundational issues. For example, in the return to equilibrium gravitational waves radiate past infinity \cite{Misner:1974qy} and are presumably ignored. How is this to be incorporated into statistical mechanics? Will two different microstates be the same microstate after this ``cropping" process?

It is widely expected that quantum gravity will be required for the correct black hole entropy derivation from statistical mechanics. However, black hole statistical mechanics may currently lack an equivalent of the Rayleigh Jeans catastrophe - that is, a classical calculation that proves to be wrong but is a helpful guide to the correct quantum description. The classical calculation is usually assumed to result in zero black hole entropy because of the black hole uniqueness theorems, but the relevant phase space involves more than a single solution when the rotational or angular velocity ensembles are used. Taking this a step further, it would be interesting, although extremely difficult, to calculate the rotational or angular velocity ensemble partition functions, using the Hamiltonian of general relativity, over a phase space representing all solutions to Einstein's field equations, including non-stationary ones. After all, stationarity is a property of the ensemble, not the individual microstates. As the microstates would then include metrics with and without horizons, such a calculation may be of relevance to the Mathur conjecture that black hole microstates in string theory have no horizons \cite{Mathur:2005zp}.

\bibliographystyle{plain}
\bibliography{ut-thesis}

\end{document}